\documentclass{iau}
\usepackage{graphicx,natbib}
 
\newcommand{\apj}{ApJ}           
\newcommand{\mnras}{MNRAS}       

\newcommand{\aap}{A\&A}

\title{News from the  isolated ellipticals NGC 5812, NGC 7507, and NGC 7796 }

\author[]{Tom Richtler$^1$, Ricardo Salinas$^2$, Richard Lane$^1$, Michael Hilker$^3$, Juan Pablo Caso$^4$ \and Lilia P.  Bassino$^4$ }

\affiliation{$^1$Departamento de Astronom\'{\i}a, Universidad de Concepci\'on, Chile  \\
$^2$Department of Physics and Astronomy, Michigan State University, East Lansing, Michigan, USA \\
$^3$European Southern Observatory, Garching, Germany\\
$^4$Universidad Nacional de La Plata and IALP (CONICET-UNLP), La Plata, Argentina\\ 

}

\pubyear{2014}
\volume{309}
\jname{Galaxies in 3D across the Universe}
\editors{B. L. Ziegler, F. Combes, H. Dannerbauer, M. Verdugo, eds.}

\begin{document}

\maketitle

\begin{abstract}
We report on ongoing photometric and spectroscopic work  on a sample of isolated elliptical
galaxies. We investigate their globular cluster systems, and use the kinematics of globular clusters and the integrated galaxy light to constrain their dark halos, which are not found in the cases of NGC 5812 and NGC 7507.    

\keywords{galaxies: elliptical and lenticular, cD - galaxies: evolution - galaxies: formation - galaxies: halo}
\end{abstract}

\firstsection
\section{Introduction}
\noindent
Isolated galaxies are expected to evolve differently from galaxies in groups or clusters. This should be manifest particularly in the structures of
their visible and dark halos, where globular clusters (GCs)  can be tracers for both components. 
 We collected photometric data, using VLT, Gemini South, and 4m Blanco telescope,
of a dozen isolated ellipticals. Besides characterising their GC systems, we want to identify clusters suitable for the use as dynamical
tracers for testing 
cosmological simulations \citep{niemi10}.

\section{NGC 5812 - alone in the dark?}
\noindent
 NGC 5812 (distance 28 Mpc) is a galaxy with a strong intermediate-age stellar component. The accompanying dwarf galaxy at a projected distance of about 20 kpc shows an
extended tidal tail. Photometric data come from the 4m-Blanco telescope.  \citet{lane13} describes its morphology, photometric properties and cluster system, which is rather poor.
We obtained mask spectroscopy with GMOS/Gemini-South (GS-2013B-Q51; PI: Richtler) and got radial velocities for 25 GCs out to a projected radius of  20 kpc. The brightest
object
has about $M_R \approx -12$, confirming the existence of clusters of this brightness as  statistically predicted
in the CMD shown by  Lane et al. One cluster has a strongly deviating velocity of 1440 km/s (systemic velocity 1970 km/s), while 24 objects show a tight distribution around the systemic velocity
of NGC 5812  with a dispersion of only 95 km/s. This low value is puzzling.  A sample of 24 GCs should already permit a reliable estimation of the total velocity
dispersion of the cluster system. The central velocity dispersion of 200 km/s (HyperLeda) requires with our photometric model a stellar $M/L_R$-value of about
3.5 under isotropy. While we await a more complete dynamical analysis, we anticipate that it  will be difficult to reconcile the low velocity dispersion with the existence  of a massive dark halo. 
NGC 5812 thus seems to be another candidate among isolated ellipticals for having less dark matter than expected.

\section{NGC 7507 - little dark matter}
\noindent
 NGC 7507 (distance 27 Mpc) belongs to  the galaxies with a Keplerian decline of the projected velocity
 dispersion \citep{salinas12}. It also has a quite poor GC system \citep{caso13} and we could
 not use GCs  as dynamical tracers. 
The kinematic data within 1 arcmin stem from long-slit observations.  By deep mask spectroscopy, obtained with Gemini/GMOS (GS-2009B-Q84; PI: Salinas) we extended the radius with measured velocity dispersions out to 130 arcsec 
\citep{lane14}. The adjacent 
regions agree excellently, but at 70 arcsec, we observe a bump in the velocity dispersion profile, which according to
\citet{schauer14} might be the relic of a former merging event. At larger radii, the velocity dispersion profile is
again well described by the stellar mass alone. Radial anisotropy and the inclusion of rotation may help to
accommodate some dark matter.

\section{NGC 7796 - an isolated cluster elliptical}
\noindent
Deep imaging with  VLT/VIMOS in B and R (89.B-457; PI: Salinas) builds the database for our investigation of the GC system of NGC 7796 (distance 50 Mpc), which is a massive old galaxy without striking substructure.
  However, a companion dwarf galaxy shows tidal tails and multiple nuclei/star cluste bluer than their parent galaxy. In contrast to many other isolated ellipticals, the GC system of NGC 7796 with estimated 2000 members
  rivals  some cluster ellipticals. We derive a specific frequency of $S_N = 2.6\pm0.5$. We see the familiar
  bimodal distribution in B-R  which characterises an old cluster system. The density profiles of blue and red GCs
  agree within the uncertainties. Therefore the growth of the halo  probably did not happened through later accretion
  of  dwarf galaxies, as expected  for an isolated elliptical. More insight will come from comparing the
  kinematics of  metal-poor and metal-rich GCs. 
  The kinematical  literature data for the galaxy centre result in $M/L_R$=6.5 for the stellar component under isotropy, consistent
with an old metal-rich population, but do not
  permit solid statements regarding the dark matter content. Assuming MOND, $M/L_R$ becomes 6.2, and  the X-ray data of \citet{osullivan07}  are consistent
  with the MONDian prediction.  
 In comparison with NGC 7507 and NGC5812,  NGC7796 may indicate that besides accretion, the epoch of star formation determines the richness of a GC system. 
    


\section*{Acknowledgements}

\noindent
TR acknowledges support from  FONDECYT project Nr.\,1100620,
 the BASAL Centro de Astrof\'isica y Tecnolog\'ias Afines (CATA) PFB-06/2007, and 
a  visitorship at ESO/Garching.

\end{document}